\documentstyle[12pt]{article}
\begin{document}

\newtheorem{theorem}{Proposition}

\begin{center}
{\Large{\bf Helicity current as a symplectic dilation}} \\

\vspace{13mm}
{\large Hasan G\"{u}mral   \\
\vspace{5mm}
Feza G\"{u}rsey Institute  \\
P.O. Box 6, 81220 \c{C}engelk\"oy-\.Istanbul, Turkey} \\
\vspace{2mm}
hasan@gursey.gov.tr \\
\vspace{7mm}
\today
\end{center}

\vspace{1cm}

\section*{Abstract}

A formal symplectic structure on $R \times M$ is constructed
for the unsteady flow of an incompressible viscous fluid
on a three dimensional domain $M$.
The evolution equation for the helicity density is expressed
via the divergence of the associated Liouville vector field
that generates symplectic dilation.
For an inviscid fluid this equation reduces to a conservation law.
As an application the symplectic dilation is used to generate
Hamiltonian automorphisms of the symplectic structure which
are then related to the symmetries of the velocity field.

\newpage

The helicity which is first discovered in \cite{skr66}
has been recognized to be an important ingradient of the
problem of relationship between invariants of fluid motion
and the topological structure of the vorticity field
\cite{FRI}-\cite{ark92}.
For three-dimensional flows its ergodic and topological
interpretations were introduced and investigated in
\cite{mof69}-\cite{gik92}.
It has also been studied in the context of Noether theorems 
\cite{mor61}-\cite{pad98}.
Kinematical aspects of helicity invariants in connection
with the particle relabelling
symmetries were discussed in \cite{hg99}.

In this work, we shall show that there is also a dynamical
content of the helicity density in the sense that the information
contained in the Eulerian dynamical equations can be represented
in the framework of symplectic geometry by a current vector field
governing the dynamics of helicity. More precisely,
starting from the Navier-Stokes equations of incompressible fluids
we shall construct helicity four-vector whose divergence will define
the time-evolution of helicity density.
The dynamical properties of the fluid, such as viscosity, are
implicit in this vector field.
The evolution equation for the helicity density reduces to a conservation
law for inviscid Euler flows.
For fluid dynamical content of this work we shall refer to \cite{FRI}
and the necessary mathematical background can be found in
\cite{via}-\cite{LM}.

The Navier-Stokes equation for a viscous incompressible fluid in a
bounded domain $M \subset R^3$ is
\begin{equation}
  {\partial {\bf v} \over \partial t} +  {\bf v} \cdot \nabla {\bf v}
  = - \nabla p +\nu \nabla^2 {\bf v} \label{euler}
\end{equation}
where ${\bf v}$ is the divergence-free velocity field tangent to
the boundary of $M$, $p$ is the pressure per unit density and $\nu$
is the kinematic viscosity.
The identity ${\bf v} \cdot \nabla {\bf v}  = 
{1 \over 2} \nabla | {\bf v}|^{2} - {\bf v} \times (\nabla \times {\bf v} )$
can be used to bring the equation (\ref{euler}) into the form
\begin{equation}
  {\partial {\bf v} \over \partial t} -
  {\bf v} \times (\nabla \times {\bf v}) = \nu \nabla^2 {\bf v}
  - \nabla (p + {1 \over 2} v^{2} )   \label{reuler}
\end{equation}
and in terms of the vorticity field ${\bf w} \equiv \nabla \times {\bf v}$
this gives
\begin{equation}
    {\partial {\bf w} \over \partial t} - 
    \nabla \times ( {\bf v} \times {\bf w} )
     = \nu \nabla^2 {\bf w}  \;.    \label{weq}
\end{equation}
For a fluid with a potential $\varphi$ and velocity field ${\bf v}$
the densities
\begin{equation}
   {\cal H} = {1 \over 2} {\bf v} \cdot \nabla \times {\bf v} \;,\;\;\;
   {\cal H}_w = {1 \over 2} {\bf w} \cdot \nabla \times {\bf w} \;,\;\;\;
       q=  {\bf w} \cdot \nabla \varphi =w(\varphi) 
\end{equation}
will be called helicity, vortical helicity and potential vorticity,
respectively.
\begin{theorem}
For a velocity field satisfying Eqs.(\ref{euler}) and for
$q \neq 2 \nu {\cal H}_w$ the two-form
\begin{equation}
  \Omega_{\nu} = - (\nabla \varphi + {\bf v} \times {\bf w} 
        - \nu \nabla \times {\bf w} ) \cdot d{\bf x} \wedge dt 
        + {\bf w} \cdot (d{\bf x} \wedge d{\bf x})    \label{sympd}
\end{equation}
is symplectic on $I \times M$ where $I$ is an open interval in $R$.
Moreover, it is exact, $\Omega_{\nu} = d \theta$ with the Liouville
(or canonical) one-form
\begin{equation}
  \theta = - ( \varphi + p + {1 \over 2} v^{2} )
   \, dt + {\bf v} \cdot d{\bf x}     \label{cone2}
\end{equation}
which is independent of the viscosity $\nu$.
\end{theorem}
{\bf Proof:} $\Omega_{\nu}$ is closed by Eq. (\ref{weq}) and the
divergence-free property of the vorticity field. The non-degeneracy
follows from the recognition that the density in the symplectic
volume $\Omega_{\nu} \wedge \Omega_{\nu} /2$ is the function
$q-2\nu {\cal H}_w$ which is assumed to be non-zero. The exactness
can be verified using Eq. (\ref{reuler}). $\bullet$

For an arbitrary smooth function $f$ of $(t,x)$ the unique Hamiltonian
vector field $X_f$ defined by the symplectic two-form (\ref{sympd}) via
$i(X_f)(\Omega_{\nu})= -df$ is given by
\begin{equation}
  X_{f}={1 \over q-2 \nu {\cal H}_{w} }
          [ - w(f) ({\partial \over \partial t}+v)
  +{df \over dt} w + (( \nabla \varphi -\nu \nabla \times {\bf w})
      \times \nabla f) \cdot \nabla ]  \;.
\end{equation}
Here, $d/dt$ denotes the convective derivative $\partial_t+ {\bf v}
\cdot \nabla$ which, viewed as a vector field on $I \times M$,
is not Hamiltonian. In fact, with the notation $v \equiv {\bf v} \cdot
\nabla$, one can check that the one-form
\begin{equation}
  i(\partial_t+v)(\Omega_{\nu})=(\nabla \varphi - \nu \nabla \times {\bf w})
    \cdot  (d {\bf x} - {\bf v} dt)
\end{equation}
is not closed and hence $\partial_t+v$ is not even locally Hamiltonian.

Next proposition describes invariantly the connection between
the symplectic structure (\ref{sympd}) and the the helicity density.
\begin{theorem}     \label{ee}
The identity
\begin{equation}
    d (\theta \wedge \Omega_{\nu} ) -
       \Omega_{\nu} \wedge \Omega_{\nu} \equiv 0           \label{helc}
\end{equation}
gives the equation
\begin{equation}
 { \partial {\cal H} \over \partial t} + \nabla \cdot
 (  {\cal H} {\bf v} + {1 \over 2} ( p - {1 \over 2} {\bf v}^2 ) {\bf w} )
 = {\nu \over 2} {\bf v} \cdot \nabla^2 {\bf w} - \nu {\cal H}_w  \label{hel}
\end{equation}
for the evolution of helicity density.
\end{theorem}
{\bf Proof:}
We have $\Omega_{\nu} \wedge \Omega_{\nu} = -2(q-2\nu {\cal H}_w)
dx \wedge dy \wedge dz \wedge dt$ and we compute
\begin{equation}
   \nabla \cdot [ \varphi {\bf w} +  {\bf v} \times ( \nabla \varphi -
        \nu \nabla \times {\bf w}) ]  =
        2q - 2 \nu {\cal H}_w  - \nu {\bf v} \cdot \nabla^2 {\bf w}
\end{equation}
for the derivative of certain terms in the expression
\begin{eqnarray}
   & &  \theta_{\nu} \wedge \Omega_{\nu}  = 
  2{\cal H} \, dx \wedge dy \wedge dz - \nonumber \\
       & & \;\;\;\; [ ( \varphi +p - {1 \over 2} v^2 ) {\bf w}
        + 2{\cal H} {\bf v}
        +  {\bf v} \times ( \nabla \varphi -
        \nu \nabla \times {\bf w}) ]  \cdot
       d{\bf x} \wedge d{\bf x} \wedge dt \;\;\;
\end{eqnarray}
for the three-form. Putting them together in the identity (\ref{helc})
we obtain Eq. (\ref{hel}). Upon integration, the term
$\nu  {\bf v} \cdot \nabla^2 {\bf w} /2$ in Eq. (\ref{hel})
gives the integral of $- \nu {\cal H}_w$ and one obtains the
usual expression for the time change of total helicity
as given in, for example, Ref. \cite{FRI}. $\bullet$

Note that the helicity flux in Eq. (\ref{hel}) is independent of
the function $\varphi$ which we have introduced by hand to make the
symplectic form non-degenerate.

Using the invariant description (\ref{helc}) of the evolution of
helicity density, we shall introduce a current vector $J_{\nu}$ and
show that it is an infinitesimal symplectic dilation of $\Omega_{\nu}$.
$J_{\nu}$ will be defined as the one-dimensional kernel of the three-form
$\theta \wedge \Omega_{\nu}$.
Since the symplectic two-form is nondegenerate, it can be obtained as
the unique solution of
\begin{equation}
      i(J_{\nu})(\Omega_{\nu} \wedge \Omega_{\nu} /2)=
                  \theta \wedge \Omega_{\nu}  \;,         \label{defj}
\end{equation}
that is, as the dual of the three-form
$\theta \wedge \Omega_{\nu}$ with respect to the symplectic volume.
We find
\begin{equation}
   J_{\nu} = {1 \over q-2 \nu {\cal H}_{w} } [  2{\cal H} (\partial_t + v)
  + ( \varphi +p - {1 \over 2} v^2 ) w  
    +  {\bf v} \times ( \nabla \varphi -
        \nu \nabla \times {\bf w})   \cdot \nabla ]  
\end{equation}
as the expression for the helicity current.

\begin{theorem} $J_{\nu}$ is a vector field of divergence $2$ with
respect to the symplectic volume and it is an infinitesimal
symplectic dilation for $\Omega_{\nu}$.
The evolution of helicity density ${\cal H}$ can be described by
the identity
\begin{equation}
    div_{\Omega_{\nu}}(J_{\nu}) -2 \equiv 0  \;.          \label{divj}
\end{equation}
\end{theorem}
{\bf Proof:}
The exterior derivative of Eq. (\ref{defj}) gives
\begin{eqnarray}
  d i(J_{\nu})(\Omega_{\nu} \wedge \Omega_{\nu} /2)&=&
 {\cal L}_{J_{\nu}}( \Omega_{\nu}\wedge \Omega_{\nu}/2) \equiv
   div_{\Omega_{\nu}}(J_{\nu}) \, \Omega_{\nu}\wedge \Omega_{\nu}/2   \\
     &=& d (\theta \wedge \Omega_{\nu} ) 
       \; = \; \Omega_{\nu}\wedge \Omega_{\nu}
\end{eqnarray}
where we used the identity ${\cal L}_{J} = i(J) \circ d + d \circ i(J)$
in the first equality and the second equality is the definition of the
divergence. We see that $J_{\nu}$ is a vector field whose divergence is $2$.
From the last equality, we conclude that
the equation (\ref{divj}) is equivalent to Eq. (\ref{hel}) describing
the evolution of helicity density.
$J_{\nu}$ is the unique vector field satisfying
\begin{equation}
          i(J_{\nu})(\Omega_{\nu}) = \theta            \label{teta}
\end{equation}
and it follows from this that $J_{\nu}$ fulfills the condition
\begin{equation}
     {\cal L}_{J_{\nu}} (\Omega_{\nu}) = d i(J_{\nu}) (\Omega_{\nu})
           = d \theta = \Omega_{\nu}             \label{ome}
\end{equation}
of being an infinitesimal symplectic dilation for $\Omega_{\nu}$
\cite{alan}. $J_{\nu}$ is also called to be the Liouville vector
field of $\Omega_{\nu}$ \cite{LM}.  $\bullet$

We observed that the local existence of a Hamiltonian function
for $\partial_t+v$ is being prevented by the viscosity term \cite{hg97}.
Moreover, the viscosity term causes the helicity not to be conserved.
We shall now show that, for the case of inviscid incompressible fluids
described by the Euler equation, namely Eq. (\ref{euler}) with $\nu=0$,
$\partial_t+v$ is Hamiltonian and that the helicity density ${\cal H}$
is conserved. To this end, we assume that the scalar
field $\varphi$ is advected by the fluid motion
\begin{equation}
    {\partial \varphi \over \partial t} +
     {\bf v} \cdot \nabla \varphi =0     \label{he}
\end{equation}
and that the potential vorticity $ q \neq 0$.
\begin{theorem} \cite{hg97}
Let $v$ and $\varphi$ satisfy
Eq. (\ref{euler}) with $\nu=0$ and Eq.(\ref{he}), respectively.
Then, the suspended velocity field
$\partial_{t}+v$ on $I \times M$ and $q^{-1}w$ are Hamiltonian
vector fields for the exact symplectic two-form
\begin{equation}
  \Omega_{0} = - (\nabla \varphi + {\bf v} \times {\bf w}) 
          \cdot d{\bf x} \wedge dt + {\bf w} \cdot 
             (d{\bf x} \wedge d{\bf x}) = d \theta   \label{sympo}    
\end{equation}
with the Hamiltonian functions $\varphi$ and $t$, respectively. 
The evolution equation (\ref{hel}) reduces to
the conservation law in divergence form for the helicity density.
\end{theorem}
{\bf Proof:}
Using Eq. (\ref{he}) $\partial_{t}+v$ can be written in Hamiltonian
form $i(\partial_t+v)(\Omega_0)= - d\varphi$. More generally,
the Hamiltonian vector field with the symplectic
two-form (\ref{sympo}) for an arbitrary function $f$ on $I \times M$
is given by
\begin{equation}
  X_{f}={1 \over q } [ -  w(f)
      ({\partial \over \partial t}+v) +{df \over dt} w
      + (\nabla \varphi \times \nabla f) \cdot \nabla ]
\end{equation}
which clearly reduces to $\partial_t+v$ for $f=\varphi$ and to
$q^{-1}w$ for $f=t$.
The conservation of helicity density is obvious. $\bullet$

For the inviscid flow of the Euler equation the helicity
current takes the form
\begin{equation}
   J_{0} ={1 \over q} [  2{\cal H} (\partial_t + v)
  + ( \varphi +p - {1 \over 2} v^2 ) w  
    +  {\bf v} \times  \nabla \varphi   \cdot \nabla ]  
\end{equation}
while the canonical one-form remains to be the same.
That means, the difference between the dynamics of fluid
motion with $\nu =0$ and $\nu \neq 0$ is contained in the
helicity current.
Thus, the dynamical content
of the helicity is encoded in its current and this, in turn, is connected
with the symplectic structure on $I \times M$ which was
constructed as a consequence of the Eulerian dynamical equations.

The realization of dynamics of fluid motion in the symplectic
framework is useful in the study of the geometry of the motion
on $M$ and of the hypersurfaces in $I \times M$ defined by
the time-dependent Lagrangian invariants, that is, the
invariants of the velocity field.
The present framework also provides geometric tools for the
investigation of scaling properties of the fluid motion 
because the action by the Lie derivative of helicity
current on tensorial objects corresponds to infinitesimal scaling
transformations \cite{LM}.
Leaving the discussions of these issues elsewhere,
we shall conclude this work with an application to the symmetry
structure of the velocity field which is also related to the results
presented in \cite{hg98}.
\begin{theorem}
Let $X_f$ be a Hamiltonian vector field for $\Omega_{\nu}$. Then,
the vector fields $({\cal L}_{J_{\nu}})^k(X_f), \; k=0,1,2,...$
are infinitesimal Hamiltonian automorphisms of $\Omega_{\nu}$.
\end{theorem}
{\bf Proof:}
The symplectic two-form is invariant under the flows of 
Hamiltonian vector fields because
${\cal L}_{X_f}(\Omega_{\nu}) = d i(X_f)(\Omega_{\nu}) = d^2f \equiv 0$   
where we used the identity ${\cal L}_{X}=i(X) \circ d+d \circ i(X)$
for the Lie derivative, $d \Omega_{\nu} =0$ and the Hamilton's equations
$i(X_f)(\Omega_{\nu}) = -df$.
It then follows from the identity
\begin{equation}
{\cal L}_{[ J_{\nu},X_f ]}=
  {\cal L}_{J_{\nu}} \circ {\cal L}_{X_f} -
  {\cal L}_{X_f}  \circ {\cal L}_{J_{\nu}}    \label{iden}
\end{equation}
evaluated on $\Omega_{\nu}$ that $[ J_{\nu},X_f ]$
also leaves $\Omega_{\nu}$ invariant.
Replacing $X_f$ with $[ J_{\nu},X_f ]$
in Eq. (\ref{iden}) we see that one can generate an
infinite hierarchy of invariants of the symplectic two-form
$\Omega_{\nu}$. To see that these are Hamiltonian vector fields
we compute
\begin{eqnarray}
  i([ J_{\nu},X_f  ])(\Omega_{\nu}) &=&
   {\cal L}_{J_{\nu}} ( i(X_f) (\Omega_{\nu})) -
   i(X_f) ( {\cal L}_{J_{\nu}} (\Omega_{\nu}))     \label{jx}  \\
   &=& - d ( J_{\nu}(f) -f) 
\end{eqnarray}
where we used Eq. (\ref{ome}). Thus, $[ J_{\nu},X_f  ]$ is Hamiltonian
with the function $J_{\nu}(f) -f$. By induction one can find
similarly that $({\cal L}_{J_{\nu}})^2(X_f)$ is Hamiltonian with
$(J_{\nu})^2(f)-2J_{\nu}(f)+f$ and so on.
Interchanging $J_{\nu}$ and $X_f$ in
the identity (\ref{jx}) we also obtain $i(X_f)(\theta) = J_{\nu}(f) $.
$\bullet$

In particular, we let $\nu=0$, $f=t$ so that $X_t=q^{-1}w$
and consider the infinitesimal Hamiltonian automorphisms
$({\cal L}_{J_{0}})^k(q^{-1}w), \; k=0,1,2,...$ of $\Omega_0$.
The identity (\ref{iden}) evaluated on the vector field
$\partial_t+v$ gives
\begin{equation}
{\cal L}_{[ J_{0},q^{-1}w ]}(\partial_t+v)=
 - {\cal L}_{q^{-1}w} ([ J_{0},\partial_t+v ])       \label{sym}
\end{equation}
where the vector field $[ J_{0},\partial_t+v ]$ is, by proposition (5),
Hamiltonian with the function $J_{0}(\varphi) -\varphi = p-v^2/2$.
By the Lie algebra isomorphism $[X_f,X_g]= X_{ \{ f,g \} }$
defined by the symplectic structure $\Omega_0$, the right hand
side of Eq. (\ref{sym}) is a Hamiltonian vector field with the
function
\begin{equation}
  \{ t,p-{1 \over 2}v^2 \} =  {1 \over q} w(p-{1 \over 2}v^2) \;.
\label{ff}    \end{equation}
On the level surfaces defined by the constant values of the
function (\ref{ff}) we have $[[ J_{0},q^{-1}w ], \partial_t+v]= 0$
In fact, if we restrict to the constant values of the function
$p-v^2/2$ the hierarchy of Hamiltonian automorphisms of $\Omega_0$
can be identified as the infinitesimal symmetries of the velocity field.
This can be seen by replacing
$q^{-1}w $ with $[ J_{0},q^{-1}w ]$ in Eq. (\ref{sym}).
We thus proved that
\begin{theorem}
For the Euler flow,
the hierarchy of infinitesimal Hamiltonian automorphisms
$({\cal L}_{J_{0}})^k(q^{-1}w), \; k=0,1,2,...$ of $\Omega_0$
generate infinitesimal time-dependent symmetries of the velocity
field on the level surfaces $p-v^2/2 = constant$.
\end{theorem}
As a matter of fact, the function $p-v^2/2$ is related, in Ref. \cite{pam},
to the invariance under particle relabelling symmetries of the
Lagrangian density of the variational formulation of the
Euler equation.

\end{document}